\documentstyle[preprint,aps,prl,eqsecnum,epsf]{revtex}
\begin{document}
\draft
\preprint{turbcrit6.tex}
\title{
  Critical behavior of vorticity in two-dimensional turbulence
}
\author{
  {\sc  Denis Boyer}
}
\address{ 
    {\it Departamento de F\'\i sica,
    Facultad de Ciencias F\'\i sicas y Matem\'aticas, \\
    Universidad de Chile,
    Casilla 487-3, Santiago, Chile.}
}
\date{\today}
\maketitle
\begin{abstract}
We point out some similitudes between the statistics of
high Reynolds number turbulence and critical phenomena.
An analogy is developed for two-dimensional decaying flows, 
in particular by studying
the scaling properties of the two-point vorticity correlation function
within a simple phenomenological framework. The inverse of the Reynolds
number is the analogue of the 
small parameter that separates the system from criticality.
It is possible to introduce a set of three critical exponents;
for the correlation length, the autocorrelation function and a so-called 
susceptibility, respectively. The exponents corresponding to the well-known
enstrophy cascade theory of Kraichnan and Batchelor are, remarkably,
the same as the Gaussian approximation exponents for spin models. 
The limitations of the analogy, in particular the lack of universal
scaling functions, are also discussed.
\end{abstract}
\pacs{PACS numbers: 47.27.Jv, 05.40.-a, 05.70.Jk}

\section{Introduction}

Turbulence is characterized in a generic way by self-similar spectra, a 
feature that is also found in other physical situations, such as critical 
phenomena or random walks. In real space,
this property is usually the signature of well-known
statistical mechanisms such as the
inverse power law decay of two-point correlation functions
and the non-Gaussian behavior of the relative dispersion of random variables
of the problem. Some attempts have been made to relate scaling in turbulence
with critical phenomena and field theories \cite{eyink}.
More recently, experiments have shown that the power consumption in closed 
turbulent flows and the total magnetization of spin systems at criticality
have similar, and to some extent, universal, 
probability distribution functions \cite{nature}.
The two-point velocity correlation function has been a 
quantity extensively studied in hydrodynamic turbulence, mostly under
the related form of velocity increments \cite{frisch}. 
In practice, real space experimental
and numerical analysis of eventual scaling laws is a difficult, space and
time consuming task, because the asymptotic regimes are reached very slowly
as the Reynolds number goes to infinity. On the other hand, relatively few 
studies have been devoted to the characterization of the spatial structure 
of the vorticity ($\vec{\omega}=\nabla\times{\bf u}$).
Although, in Fourier space, the enstrophy spectrum of an incompressible flow
is very simply related to the kinetic energy spectrum,
the vortex statistics and correlations in real space are not
so simply related to their velocity counterparts, but have distinct
properties. This difference is illustrated in a spectacular way in 
two-dimensional turbulence, where the enstrophy
$\langle \vec{\omega}^2({\bf x})\rangle$ cascades toward small scales 
while the kinetic energy $\langle {\bf u}^2({\bf x})\rangle$ 
follows an inverse cascade to large scales.
This is essentially due to the absence of vortex stretching in two dimensions. 
Because of this property, the enstrophy obeys a 
conservation equation, and
$\langle \vec{\omega }^2({\bf x})\rangle$ is
dissipated even in the very low viscosity limit,
contrary to the energy \cite{frisch,lesieur}.

A vortex is one of the simplest kind of structure generated by hydrodynamic
instabilities; an isolated structure is likely to produce slowly decaying flow 
fields in space (typically with $1/r$, $1/r^2$ dependence) and consequently 
long range velocity correlations. 
Physically, a flow can be structured even if the vorticity vanishes nearly
everywhere, as in the case of an assembly of point vortices in two dimensions. 
Hence, the study of vorticity distribution functions
should reveal a deeper, less intuitive degree of order in flows.
This occurs in two-dimensional turbulence, where the 
vorticity tends to concentrate in spatially extended structures, of
particle-like character. 
Indeed, numerical studies of two-dimensional decaying turbulence
have shown that the long time dynamics is dominated by a few large coherent
axisymmetric vortices that contain most of the enstrophy \cite{benzi}.
Dynamically, it is commonly believed that these structures are the result
of many vortex mergers. Such flows can be often well described by a
modified Hamiltonian, deterministic model of point vortices \cite{carnevale}. 
These rather
well isolated structures of well defined shape and size have the advantage
of simplifying the picture of the flow, because the organization of initially
numerous incoherent vortices into blobs reduces the number of relevant degrees
of freedom.

In this study, following a route similar to the analysis by Babiano {\it et al.}
\cite{babiano} of velocity structure functions, we characterize the vorticity 
structure in real space for 
two-dimensional homogeneous unforced turbulence, keeping in mind its 
possible representation in terms of assemblies of vortices. We consider
the normalized vorticity autocorrelation function, and calculate it
assuming a phenomenological self-similar
energy spectrum in the inertial range, $E(q)\sim q^{-\mu}$. In experiments
or numerical studies, $\mu$ is usually found to be around 3 \cite{soap}, 
the value predicted  by the Kraichnan \cite{kraichnan} and Batchelor 
\cite{batchelor} reference theory, hereafter referred to as KB, or
steeper ($\mu\simeq4$).

In the next section, we show that the vorticity autocorrelation function 
exhibits distinct behaviors depending on the value of the exponent $\mu$. 
In section III, we show that it is possible to draw an analogy between 
two-dimensional turbulence 
and systems of spins near criticality for values of $\mu$ larger than $1$.
The correlations can indeed be written under the form of a scaling law 
involving a diverging correlation length when the Reynolds number goes to
infinity. The critical exponents that correspond to the KB theory precisely
coincide with the Gaussian exponents of spin models or random walks. 
Conclusions including
a brief discussion of the three-dimensional case are
presented in section IV.

\section{Vorticity power spectra}

Let us consider the vorticity autocorrelation function
\begin{equation}\label{romdef}
R_{\omega}({\bf r})=\frac{\left\langle\vec{\omega }({\bf x})\cdot
\vec{\omega }({\bf x}+{\bf r})\right\rangle}
{\left\langle\vec{\omega }^2({\bf x})\right\rangle}\ ,
\end{equation}
that has been normalized to one at ${\bf r}={\bf 0}$. $R_{\omega}$
can be expressed as a function of the Fourier transform of the vorticity,
\begin{equation}
\vec{\omega }({\bf q})=\int d{\bf x}\ \vec{\omega }({\bf x})\ 
e^{-i{\bf q}.{\bf x}},
\end{equation}
through the relation
\begin{equation}\label{romvort}
R_{\omega}({\bf r})=\frac{\int d{\bf q}\ \left\langle\vec{\omega }({\bf q})
\cdot\vec{\omega }(-{\bf q})\right\rangle\ e^{i{\bf q}.{\bf r}}}
{\int d{\bf q}\ \left\langle\vec{\omega }({\bf q})\cdot
\vec{\omega }(-{\bf q})\right\rangle}\ .
\end{equation}
The vorticity power spectrum $\langle\vec{\omega }({\bf q})\cdot
\vec{\omega }(-{\bf q})\rangle$ can easily be related to the kinetic energy
spectrum $E(q)$ of the flow ${\bf u}$. In two dimensions one gets, in
the homogeneous isotropic case,
\begin{equation}
E(q)=\frac{q}{4\pi V}\left\langle{\bf u}({\bf q})\cdot{\bf u}(-{\bf q})\right
\rangle\ ,
\end{equation}
where $V$ is the volume of the flow. If the flow is incompressible, one has the
identity $\langle\vec{\omega }({\bf q})\cdot\vec{\omega }(-{\bf q})\rangle=
q^2\langle{\bf u}({\bf q})\cdot{\bf u}(-{\bf q})\rangle$, so that
\begin{equation}\label{vortEq}
\left\langle\vec{\omega }({\bf q})\cdot\vec{\omega }(-{\bf q})\right\rangle=
4\pi V\ qE(q)\ .
\end{equation}
Replacing the identity (\ref{vortEq}) in expression (\ref{romvort}) and 
integrating over the angular variable gives:
\begin{equation}\label{romvort2}
R_{\omega}(r)=\frac{\int dq\ q^2E(q)\ J_0(qr)}{\int dq\ q^2E(q)}\ ,
\end{equation}
where $J_0$ is the Bessel function of the first kind. In the following, one
assumes that the energy spectrum is self similar in the inertial range, which
extends between an integral scale $l$ at which energy is (initially) injected 
and an enstrophy dissipation scale $a$:
\begin{equation}\label{Eqin}
E(q)\sim q^{-\mu},\quad 2\pi/l<q<2\pi/a\ .
\end{equation}
The spectrum is supposed to vanish at length scales smaller than $a$. 
If the observation scale $r$ is much smaller than $l$,
one can assume in addition that $E(q)=0$ for $q<2\pi/l$. (We discuss in
section IV the modifications introduced by considering the part of the
spectrum that extends beyond the inertial range.) Note that, in the general 
case, the scale $l$, and hence the integral Reynolds number, may increase with 
time, as in freely decaying processes \cite{lesieur}. This is not a restriction
to our analysis, which
deals with instantaneous spectra. Relation (\ref{romvort2}) simply turns into:
\begin{equation}\label{rom}
R_{\omega}(r)=\left(\int_{q_l}^{q_a}dq\ q^{-(\mu-2)}J_0(qr)\right)
\left(\int_{q_l}^{q_a}dq\ q^{-(\mu-2)}\right)^{-1}\ ,
\end{equation}
where 
\begin{equation}
\ 
\begin{array}{l}
q_l=2\pi/l\\ 
q_a=2\pi/a\ .
\end{array}
\end{equation} 
Note that the normalized velocity 
autocorrelation function, $R_u(r)$, has the same expression as
Eq.(\ref{rom}), replacing $\mu$ by $(\mu+2)$. It can be seen that both the
numerator and the denominator of Eq.(\ref{rom}) diverge when 
$l\rightarrow\infty$ if $\mu\ge3$. Hence, we expect the behavior
of $R_{\omega}(r)$ for $\mu>3$ to differ from its behavior for $\mu<3$,
while the analytical form of $R_u$ does 
not undergo an abrupt change at this point.
In the following we discuss the different cases encountered.

\subsection{Case $\mu>3$}

To study the correlations in the interval $r\gg a$, we can replace the 
upper limits of the integrals of Eq.(\ref{rom}) by infinity. Thus, 
$R_{\omega}(r)$ scales as a function of $r/l$ only, and can be rewritten as
\begin{equation}\label{romg31}
R_{\omega}(r)=1-(\mu-3)q_l^{\mu-3}\int_{q_l}^{\infty}dq\ 
q^{-(\mu-2)}\left[1-J_0(qr)\right]\ ,\quad a\ll r.
\end{equation}
If $3<\mu<5$, the integral of Eq.(\ref{romg31}) is finite as $q_l$ goes to
zero. After the variable change $x=qr$, one obtains the first order expansion
in $r/l$,
\begin{equation}\label{romg32}
R_{\omega}\simeq 1-\frac{\Gamma\left(\frac{5-\mu}{2}\right)}
{\Gamma\left(\frac{\mu-1}{2}\right)}\left(\frac{\pi r}{l}\right)^{\mu-3} ,
\quad a\ll r\ll l\ ;\ 3<\mu<5\ ,
\end{equation}
where $\Gamma$ denotes the Gamma function.
If $5<\mu<7$, we use the second order expansion of the Bessel function 
$J_0(x)= 1-x^2/4+O(x^4)$, and replace in Eq.(\ref{romg31}) $[1-J_0(x)]$ by
$[1-x^2/4-J_0(x)]+x^2/4$. We obtain
\begin{equation}
R_{\omega}(r)\simeq 1-\frac{\mu-3}{4(\mu-5)}(q_lr)^2-C_{\mu}(q_lr)^{\mu-3} , 
\quad 5<\mu<7\ ,
\end{equation}
with 
\begin{equation}
C_{\mu}=(\mu-3)\int_0^{\infty}dx\ x^{-(\mu-2)}[1-x^2/4-J_0(x)]\ .
\end{equation}
The leading behavior of $R_{\omega}(r)$ is given by the $r^2$-term. It is easy
to show that this property stays valid for all $\mu>5$:
\begin{equation}\label{romg33}
R_{\omega}(r)\simeq 1-\frac{\mu-3}{4(\mu-5)}(q_lr)^2\ ,\quad a\ll r\ll l.
\end{equation}
The results displayed in Eqs.(\ref{romg32}) and (\ref{romg33}) are the 
analogues of those obtained by Babiano {\it et al.} \cite{babiano} for the
velocity structure functions (in the cases $1<\mu<3$ and $\mu>3$ respectively).
The power-law exponent of the corrections for small separation distances
does not depend on $\mu$ for $\mu>5$. The first order expansions
(\ref{romg32}) and (\ref{romg33}) show that
the characteristic length ruling the decay of $R_{\omega}$ is $q_l^{-1}$.
We conclude that the vorticity correlation length is of
order of the integral scale $l$.

\subsection{Case $3/2<\mu\le 3$}

For energy spectra less steep than $-3$, the denominator of expression 
(\ref{rom}) 
is well defined as $l$ goes to infinity, but diverges when $a$ goes to zero. 
In turn, if $3/2<\mu<3$, the numerator of Eq.(\ref{rom}) does not
necessitate infra-red nor ultra-violet cut-offs to be finite, 
provided that $r>0$. This implies that in the
limit where $l/a$ is large, $R_{\omega}(r)$ can be much smaller than one
even if $r\ll l$.

If $a\ll r\ll l$, one can set $q_l=0$ and $q_a=\infty$ in the numerator of 
Eq.(\ref{rom}). Using properties of Bessel functions \cite{luke}, one obtains:
\begin{equation}\label{roml3}
R_{\omega}(r)\simeq
\frac{\Gamma\left(\frac{5-\mu}{2}\right)}
{\Gamma\left(\frac{\mu-1}{2}\right)}\ 
\left(\frac{\pi r}{a}\right)^{-(3-\mu)}.
\end{equation}
The correlation function decays algebraically with the distance $r$.
The value $\mu=3$
separates two different scaling regimes of $R_{\omega}$. At the transition,
the numerator of Eq.(\ref{rom}) behaves like $\ln(l/a)$, and making the 
variable change $x=qa$, one obtains after a simple calculation:
\begin{equation}\label{rom31}
R_{\omega}(r)=1-\frac{1}{\ln(l/a)}\int_0^{2\pi}dx\ x^{-1}
\left[1-J_0\left(x\frac{r}{a}\right)\right] ,\quad \mu=3.
\end{equation}
When the separation distance is much larger than the dissipation scale,
the correlations decay logarithmically:
\begin{equation}\label{log}
R_{\omega}(r)\simeq\frac{\ln(l/r)-\ln\pi-0.577}{\ln(l/a)}\ , 
\quad a\ll r\ll l\ ;\ \mu=3.
\end{equation}

\subsection{Case $\mu\le 3/2$}

For values of $\mu$ lower than 3/2, both the denominator and the numerator of
Eq.(\ref{rom}) diverge when $a$ goes to zero. In order to avoid convergence
problems due to oscillations when integrating the Bessel function $J_0$ over a 
large but finite interval, it is now necessary to replace the discontinuous 
energy spectrum (\ref{Eqin}) by a continuous spectrum that models better
the effects of viscous dissipation \cite{kolmo}, for instance:
\begin{equation}\label{Eqbis}
E(q)\sim
\left\{
\begin{array}{lr}
q^{-\mu}\ e^{-q^2/q_a^2}\ &{\rm if}\ q>q_l\\
0\ &{\rm if}\ q<q_l.
\end{array}
\right.
\end{equation}
In Eq.(\ref{Eqbis}), the dissipative spectrum at large wavenumbers
$q>q_a$ is described by a Gaussian function \cite{note1}.
Setting $q_l=0$, relation (\ref{romvort2}) yields:
\begin{equation}
R_{\omega}(r)=\frac{2}{\Gamma\left(\frac{3-\mu}{2}\right)} q_a^{-(3-\mu)}
\int_0^{\infty}dq\ e^{-q^2/q_a^2} q^{2-\mu}J_0(qr)\ ,
\end{equation}
that can be recast as 
\begin{equation}\label{roml3/2}
R_{\omega}(r)=\ _1F_1\left(\frac{3-\mu}{2},1;-\frac{q_a^2r^2}{4}\right)\ ,
\end{equation}
where $_1F_1$ is a hypergeometric function. 
For $q_a r\gg 1$, and if $\mu\neq 1$, $R_{\omega}$ behaves as \cite{luke}
\begin{equation}\label{roml3/2as}
R_{\omega}(r)\simeq\frac{1}{\Gamma\left(\frac{\mu-1}{2}\right)} 
\left(\frac{\pi r}{a}\right)^{-(3-\mu)}.
\end{equation}
If $\mu>1$, the above expression is positive and does not differ from the 
behavior displayed by Eq.(\ref{roml3}). In turn, if $\mu=1$, the prefactor of 
the algebraic dependence in Eq.(\ref{roml3/2as}) vanishes and relation 
(\ref{roml3/2}) simply reduces to a short range function:
\begin{equation}\label{rom1}
R_{\omega}(r)=e^{-(\pi r/a)^2}.
\end{equation}
The energy spectrum $\mu=1$ is known to correspond to a flow composed by 
statistically independent point vortices \cite{novikov}. In that limit,
it is easy to show that the vorticity correlations reduce to
\begin{equation}\label{gas}
\left. R_{\omega}(r)\right|_{\mu=1}=
\left\{
\begin{array}{lr}
1\ &{\rm if}\ r=0\\
0\ &{\rm if}\ r>0
\end{array}
\right.
\end{equation}
Replacing $a$ by zero in Eq.(\ref{rom1}), we recover the 
distribution given by Eq.(\ref{gas}).

If $\mu<1$, vortex correlations for large separation distances $r$ are still
given by relation (\ref{roml3/2as}). However, the prefactor 
$\Gamma[(\mu-1)/2]^{-1}$ 
is now {\it negative}. This case is physically very different from 
the situation $\mu>1$, where $R_{\omega}(r)$ is always a positive, slowly
decaying function. Indeed, for $\mu<1$, $R_{\omega}(r)$ decreases for small 
$r$ and has its first zero for $r\sim a$: this is the signature of a short 
range function, such as Eq.(\ref{rom1}). We conclude that $R_{\omega}$ is 
short range and characterized by a correlation length of order $a$ in the 
range $\mu\le1$.

\section{Discussion}

The results derived in the previous section show that 
the two-point vorticity autocorrelation function can be recast under 
scaling forms, such as 
\begin{equation}\label{scaling}
R_{\omega}(r)=\left(\frac{r}{a}\right)^{-\eta}f\left(r/\xi\right)\ .
\end{equation}
The above relation defines an exponent $\eta$ and a correlation length $\xi$.
Pair correlations in systems near a critical point 
satisfy similar scaling laws, where the scaling function $f$ decays, say,
exponentially, and $\xi$ diverges as the temperature
approaches the critical temperature.
In the present context,
the vorticity has various spatial structures depending on the slope 
of the energy spectrum. 
The correlation length $\xi$ is small and equals the dissipation length $a$
when $\mu<1$, see Eq.(\ref{rom1}). The case $\mu>1$ is more interesting
because, then, the correlation length is large: $\xi=l\gg a$,
see Eqs.(\ref{romg32}) and (\ref{romg33}). 
(In the range $1<\mu<3$, if the separation distance $r$ is no 
longer small compared with $l$, it is easy to check that 
expression (\ref{roml3}) or (\ref{roml3/2as}) reduces to the
form (\ref{scaling}) with $\xi=l$.)
These two distinct regimes
are indeed driven by the behavior of the
numerator of Eq.(\ref{rom}) with respect to
the ultra-violet cut-off.
When this cut-off is necessary in order to get a finite integral, the 
correlations are short range; otherwise, they are long range.
This abrupt change
is reminiscent of a similar 
feature encountered in problems of elastic
interfaces at thermal equilibrium, 
namely the transition from flat 
to rough surfaces as the surface dimension decreases
across the value $d=2$ \cite{kardar}. In that problem,
the r.m.s of the height difference between two well-separated
points can be seen as
a correlation length.
In high dimensions (analogous here to low values of $\mu$), this difference is 
constant, of order of a small cut-off scale identified with the thickness 
of the interface (here, the dissipation scale). 
However, if $d<2$, the height difference is "macroscopic": it depends on
the size of the system, {\it i.e.} 
the separation distance between the two points.

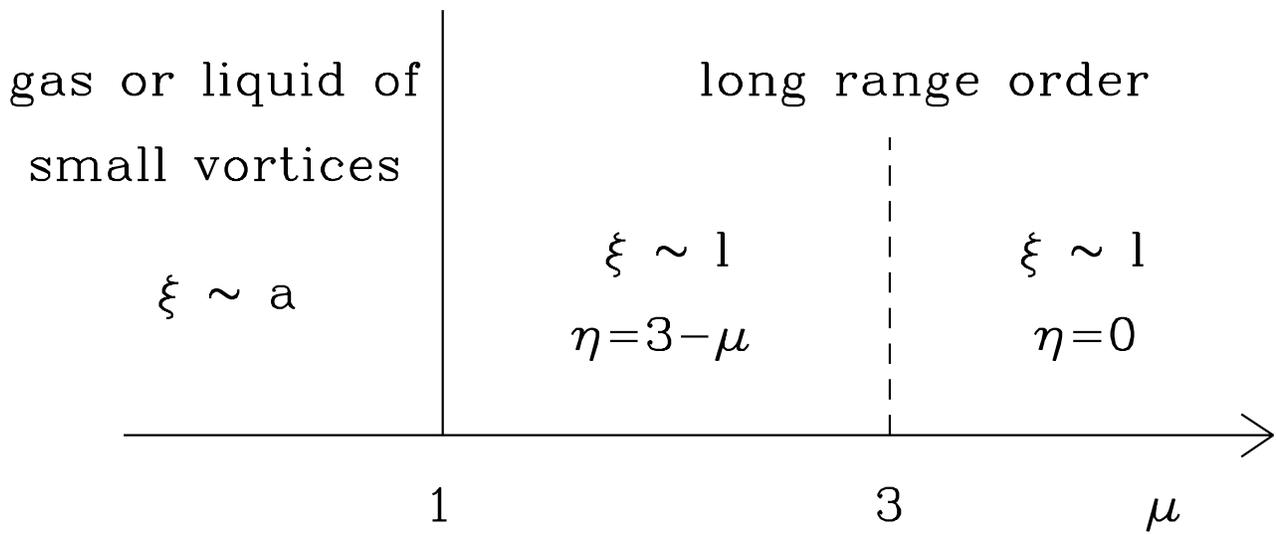
\begin{figure}
  \caption{
       Scaling forms of the vorticity autocorrelation function in the space of
       the parameter $\mu$. If the slope of the energy spectrum is lower than
       $1$, $R_{\omega}$ is a short range function. In turn, when
       $\mu>1$, the correlation length is of order of the integral scale
       $l$. Note that for $\mu>3$, the critical exponent $\eta$ is identically 
       zero.
  }
  \label{diagramme}
\end{figure}

The main scaling properties are summarized on the diagram of 
Figure \ref{diagramme}. 
Flows with short range order can be pictured for instance as a gas, or liquid, 
of nearly point-like vortices, with 
radii of order of the smallest characteristic scale,
the dissipation scale $a$. Using the language of spin 
models, these structures are also analogous to high temperature states, 
where $a$ would correspond to the lattice spacing. 
The part of the diagram where $\mu$ is larger than $1$ corresponds to
vorticity distributions with long range spatial order.
In the following, we focus on this region and develop
an analogy with critical phenomena.

\subsection{Critical exponents}

An analogy between a physical problem and critical phenomena requires that 
some of its statistical properties take a simple scaling form described 
with a few relevant parameters in a particular asymptotic limit\cite{domb}. 
This can be done, for instance, in the context of random walks, where the
analogue of the spin-spin correlation is the probability of presence (or the 
end-to-end distance distribution), and the analogue of the temperature
difference from the critical temperature is the inverse of the time (or
length of the walk) \cite{degennes,bouchaud}.

In turbulence,
the integral Reynolds number of a flow is defined as $Re=vl/\nu_0$, where $v$
is the r.m.s. of the velocity and $\nu_0$ the kinematic viscosity. 
The ratio $(l/a)^d$, where $d$ is the space dimension, 
is often seen
as the number of degrees of freedom of the flow \cite{lesieur}.
Phenomenological theories involving dimensional arguments 
\cite{kraichnan,batchelor} usually lead to power
law behaviors of $l/a$ with $Re$, say 
\begin{equation}\label{la}
l\sim a\ Re^{\nu}\ . 
\end{equation}
Since $l$ is the correlation length in the region
$\mu>1$ (see Figure \ref{diagramme}), $\nu$ can be understood as
a correlation critical exponent.
Indeed, as the Reynolds number goes 
to infinity, the correlation length $\xi=l$ diverges in units of $a$, and
the small parameter that separates the system from criticality is the inverse 
of the Reynolds number.

The correlation function exponent $\eta$ of the scaling form (\ref{scaling}) 
defines a second critical exponent. $\eta=3-\mu$ in the range $1<\mu<3$, 
and $\eta=0$ for $\mu>3$.
The transition ($\mu=3$) where correlations no longer decay algebraically
at infinite Reynolds number coincides with the spectrum of the KB theory.
At this particular point, $R_{\omega}$ decays logarithmically
as shown by Eq.(\ref{log}).
Table (\ref{table}) shows the main connections with critical phenomena,
for instance with a system of Ising's spins 
at temperature slightly above $T_c$.
However, the correspondences are purely
statistical: considering few body problems, it is obvious that vortex 
interaction and dynamics \cite{aref} can not be compared with spin interaction.
Concerning this last point, the comparison with 
spin systems is much more precise in the case of a superfluid \cite{sup}.

Different theories of turbulence will in general lead to different values 
of the exponents $\nu$ and $\eta$.
In phenomenological theories, conservation laws and
other arguments (like dimensional arguments) determine
the exponent of the energy spectrum, as well as prefactors depending
on dissipation rates. Since the size of the inertial range
must be consistent with the dissipation rates, once the shape of the spectrum 
is known and the energy injection scale is fixed, the dissipation scale $a$ 
(and consequently the value of the exponent $\nu$ in relation (\ref{la})) is 
unique. Hence, the critical exponents
$\eta$ and $\nu$ are generally not independent. 

The KB
approach to two-dimensional turbulence assumes that the enstrophy flux through
wave number $q$ is independent of $q$ at large $q$; a dimensional argument
leads to $\mu=3$ and $(l/a)^2\sim Re$ \cite{lesieur}.
The critical exponents associated to the KB theory are hence $\eta=0$ 
and $\nu=1/2$. These exponents are, remarkably, the same as the exponents of 
the Gaussian approximation in spin models, or the Brownian motion exponents 
(see \cite{bouchaud}). Although the meaning of this coincidence remains 
unclear, first, it can be noted that
a mean field approximation in the present fluid dynamics context
corresponds to the fact of neglecting the fluctuations of the enstrophy 
dissipation rate during the enstrophy cascade.
Secondly, the KB theory is based on two conservation equations: for
the kinetic energy $\langle\bf{u}^2\rangle$ and for the enstrophy 
$\langle\omega^2\rangle$. Such a description is incomplete:
indeed, in two dimensions the vorticity
follows the fluid motion in the inviscid limit, 
hence any continuous functional 
$\langle F(\omega)\rangle$ 
is invariant with time, or must satisfy a conservation equation 
in the slightly dissipative case.
Assuming the Gaussian approximation, $\langle\omega^{2n+1}\rangle=0$ and
$\langle\omega^{2n}\rangle=\langle\omega^2\rangle^n$ for any integer $n$,
the conservation of $\langle\omega^2\rangle$ implies
the conservation of any functional $F(\omega)=\sum_n a_n \omega^n$.
Hence, the KB theory ressembles a Gaussian approximation, by focusing only
on the second moment of the vorticity.

\subsection{Lack of universality}

Although microscopic details (here the dissipation processes at $q>q_a$)
disappear in the behavior of $R_{\omega}(r)$ for large separation distances, 
the analogy with critical phenomena presented in this section encounters 
a limitation when looking more closely at the scaling function $f$ in 
Eq.(\ref{scaling}). 
In the case of turbulence, $f$ does not
rigorously converge in distinct "universality classes", since it does not 
depend on the self similar energy spectrum of the inertial range only. 
The scaling function $f$ essentially depends on the
shape of the energy spectrum at low wave numbers. 
In general, the integral scale is not the 
largest scale in turbulent flows and coherent structures may emerge from
external constraints: the wave numbers lower than $2\pi/l$ contain 
a non-negligible part of the energy and this part of the spectrum is driven by
"non-universal" conditions like interactions between large eddies, depending
on the system size and boundary conditions. Hence, 
only short distance scaling
is generic in turbulence, as already noted by Eyink {\it et al.} \cite{eyink}.

Let us consider the modified spectrum:
\begin{equation}\label{spec2}
E(q)\sim\left\{ 
\begin{array}{ll}
q_l^{-(\mu_0+\mu)} q^{\mu_0}\ &\quad 0<q<q_l\\
q^{-\mu}\ &\quad q_l<q<q_a.
\end{array}
\right.
\end{equation}
In two dimensions, the exponent $\mu_0=1$ corresponds to an equipartition 
kinetic energy spectrum at low wave numbers; the value $\mu_0=3$ has also 
been proposed for decaying turbulence
\cite{lesieurherr}. It can be easily shown that in the range $r\ll l$, 
Eqs.(\ref{roml3})
and (\ref{roml3/2}) remain unchanged with the spectrum given by 
Eq.(\ref{spec2}). However, if one wishes to compute the scaling function,
in the case $\mu>3$ for instance,
Eq.(\ref{romg31}) turns into
\begin{eqnarray}
R_{\omega}(r)=f\left(\frac{r}{l}\right)&=&
1-\left(\frac{1}{\mu_0+3}+\frac{1}{\mu-3}\right)^{-1}q_l^{\mu-3}
\int_{q_l}^{\infty}dq \ q^{-(\mu-2)}[1-J_0(qr)]\ ,\\
&&\quad r\ll l ;\ \mu>3.\nonumber
\end{eqnarray}
Hence, the differences introduced by the modified spectrum in the scaling 
function $f(u)$, defined by Eq.(\ref{scaling}), already appear at the first 
order expansion in $u$: the prefactor is also a function of $\mu_0$. 
As $u$ grows, the decay of $f(u)$ depends more strongly
on the "non-universal" part. 

However, with $E(q)$ given by Eq.(\ref{spec2}), the leading behavior of 
$f(u)$ for $u\gg1$ is generically exponential. A famous example is provided
by the Ornstein-Zernike (OZ) approximation for critical phenomena,
where the spin-spin correlation function reads $\langle S(0)S(r)\rangle\propto 
\int d^2{\bf k}\ 
\exp(i{\bf k}.{\bf r})/(k^2+\xi^{-2})\sim r^{-1/2}\exp(-r/\xi)$ for $r\gg\xi$
\cite{stanley}, corresponding in the present description to $\mu_0=-1$ and 
$\mu=3$.
Recent experimental results on 
two-dimensional decaying turbulence in soap films have shown clear evidence
of spectra with both power-law and inverse power-law regimes, with respective
slopes close to the 
values $\mu_0=1.5$ and $\mu=3$ \cite{soap}.
Analytical calculations can be performed in the equipartition case 
$\mu_0=1$ and $\mu=3$:
let us rewrite the energy spectrum under the form $E(q)\sim q/(q^4+\xi^{-4})$ 
with $\xi=l$. The vorticity correlations read:
\begin{equation}\label{casigauss}
R_{\omega}(r)\propto\int_0^{\infty}dq\ q^3\frac{J_0(qr)}{q^{4}+\xi^{-4}}\ .
\end{equation}
Differing from the OZ result,
the asymptotic limit of Eq.(\ref{casigauss}) exhibits oscillations \cite{grad}:
\begin{equation}\label{fugg1}
f(u)\propto \frac{1}{u^{1/2}}\ \exp(-u/\sqrt{2})\ \cos(u/\sqrt{2}+\pi/8)\ ,
\quad u\gg1.
\end{equation}

We now outline a physical interpretation of the scaling function $f$.
As mentioned before,
some two-dimensional decaying turbulent flows with a
spectrum given by Eq.(\ref{spec2}) are well
pictured by an assembly of well separated
vortex blobs with particle-like character.
This has been, for instance, very clearly observed in some recent 
numerical simulations on homogeneous Rossby wave turbulence \cite{jap}.
Let us assume that these vortices are
all of size $\xi$ and with the same circulation modulus.
The vorticity distribution inside one blob is thus
described by the short distance behavior of the vorticity correlation 
function, see e.g. Eq.(\ref{log}). In turn, the long distance behavior of 
$R_{\omega}$ accounts for the weak correlations between different vortices. 
Following a route similar to the one presented in ref.\cite{jchemphys}, 
it is easy to show from the definition (\ref{romdef}) that,
if there are the same number of blobs with positive and negative circulation 
so that the flow has no net overall rotation,
\begin{equation}
R_{\omega}(r)\propto \frac{1}{2}[g_l(r)-g_u(r)]\ ,\quad r>\xi.
\end{equation}
Using the notations employed for ionic liquids, 
$g_l$ represents the pair correlation function between two vortices of same 
circulation ("like"), and $g_u$ the pair correlation function
between two vortices with opposite circulation ("unlike").
The oscillating decaying shape of particle pair correlation functions 
is a generic feature of liquids, in particular, those that are ionic 
\cite{hansen}. The oscillations of $R_{\omega}$ obtained in a particular case 
through Eq.(\ref{fugg1}), although having a very different physical origin,
qualitatively support the idea that the structure of homogeneous
assemblies of coherent vortices can be "liquid"-like.

\subsection{Susceptibility}

Since the autocorrelation function $R_{\omega}$ is an averaged quantity, it
cannot of course provide by itself all the details on the complexity of vortex 
structures observed in turbulence, and contained in higher
moments. Yet, large coherent vortices
can be seen as a particular manifestation of long range
order up to the integral scale, similarly to magnetization
domains in systems of spins near a critical point. 
These vortices produce
large fluctuations of vorticity on scales smaller than the
integral scale, and possibly spatial intermittency. 
The fluctuations of the total vorticity 
contained inside
an observation window of area $A$ are non Gaussian (for $\mu>1$) when $A$ is
lower than the square of the integral scale. To see this, let us define the 
circulation,
\begin{equation}\label{Omega}
\Omega_A=\frac{1}{\langle{\bf\it \omega }^2\rangle^{1/2}}
\int_A d{\bf x}\cdot\vec{\omega }({\bf x})\ .
\end{equation}
The standard deviation $\sqrt{\langle\Omega_A^2\rangle}$ of $\Omega_A$ 
from its mean value
$(\langle\Omega_A\rangle=0)$ can be estimated by considering $\Omega_A$
as a sum of random correlated variables. According to the definitions
(\ref{romdef}) and (\ref{Omega}), one gets
\begin{equation}\label{bouch}
\sqrt{\langle\Omega_A^2\rangle}
\sim\left(A\ \int_A d{\bf r}\ R_{\omega}({\bf r})\right)^{1/2}\ .
\end{equation}
When $R_{\omega}$ is short range, one obtains the Gaussian behavior,
\begin{equation}
\sqrt{\langle\Omega_A^2\rangle}\sim A^{1/2}\ \quad \mu\leq 1,
\end{equation}
at any scale larger than the dissipation scale. In turn, when the correlation
length equals the integral scale $l$,
the scaling exponent is non-Gaussian (i.e larger than $1/2$) at
scales smaller than $l$.
With the help of Eqs.(\ref{scaling}) and (\ref{bouch}), we indeed find: 
\begin{equation}\label{omall2}
\sqrt{\langle\Omega_A^2\rangle}\sim A^{\alpha}\quad(A<l^2)
\end{equation}
with
\begin{equation}\label{alpha} 
\alpha=1-\eta/4=\left\{
\begin{array}{ll}
(\mu+1)/4\quad &1<\mu<3\\
1\quad &\mu>3.
\end{array}
\right. 
\end{equation}
To derive a critical relation analogous to the one for the susceptibility, 
we invoke the arguments of Bouchaud and Georges \cite{bouchaud} in their 
statistical interpretation of the general relations between critical 
exponents.
If the area $A$ of the observation window exceeds $l^2$, one can replace the 
integration domain of
Eq.(\ref{bouch}) by $l^2$; multiplying the factor $A$ by $l^2/l^2$, one gets:
\begin{equation}\label{finsiz}
\sqrt{\langle\Omega_A^2\rangle}=(A/l^2)^{1/2}\langle\Omega_{l^2}^2\rangle^{1/2}
\sim (A/l^2)^{1/2}(l^2)^\alpha\ ,\quad A\gg l^2.
\end{equation}
Hence, the system can be pictured as $A/l^2$ independent regions of size $l$
at criticality. We define the susceptibility per unit area as
\begin{equation}\label{susc}
\chi=\frac{\langle\Omega_A^2\rangle}{A}\ .
\end{equation}
When $A$ is larger than the integral scale squared, $\chi$ becomes independent 
of $A$. Taking into account the dissipation scale $a$ in the calculations, 
we find from expression (\ref{finsiz}) combined with Eqs.(\ref{la}) and
(\ref{alpha}):
\begin{equation}\label{scalchi}
\chi\sim a^2\ (1/Re)^{-\gamma}\ ,
\end{equation}
with the exponent $\gamma$ satisfying the same relation as in critical 
phenomena:
\begin{equation}
\gamma=\nu(2-\eta)\ .
\end{equation}
Relation (\ref{scalchi}) shows that the susceptibility diverges when compared
with the "microscopic" susceptibility $a^2$. Note that 
contrary to the lattice spacing
in spin models, the dissipation length $a$ may not remain constant while
other parameters of the flow are varied. In any case,
$Re$, $\langle\omega^2\rangle$ and $a$ are independent variables.

\section{Conclusion}

We have studied the two-point vorticity correlation function in 
two-dimensional turbulent flows characterized by a self-similar energy spectrum 
within an inertial range limited by two very different length scales.
If the slope is steeper than $1$, the vorticity distribution
presents some similitudes
with the magnetization in models of spins near critical point. 

On the grounds of phenomenological theories of turbulence, the analogy with 
critical phenomena is made possible
because the ratio between the integral scale (the macroscopic scale
at which energy is injected) and the dissipation scale diverges
as a power law of the Reynolds number in the large Reynolds number limit.
The divergence of susceptibility is associated here with the increasing
fluctuations of the circulation taken around a large contour, as the
Reynolds number increases.
As a surprising result, the (non-independent) critical exponents corresponding
to the Kraichnan \cite{kraichnan} and Batchelor \cite{batchelor} (KB) theory 
are the same as those of the Gaussian approximation result in spin models.  
A possible connection between these two phenomena could be as follows:
The KB theory only considers the first non-vanishing moment of the
vorticity ({\it i.e.} $\langle \omega^2\rangle$) as an inviscid
invariant of the motion, among all the infinite number of possible invariants
associated with the vorticity.
If one invokes the Gaussian approximation
$\langle\omega^{2n}\rangle=\langle\omega^2\rangle^n$ for any integer $n$,
the conservation of $\langle\omega^2\rangle$ is indeed sufficient for any 
$C^{\infty}$ functional $F(\omega)$ to be conserved.
Another particularity of the KB theory is that
the vorticity correlations decay logarithmically for distances larger than
the dissipation scale and shorter than the integral scale.
The slope of the KB energy spectrum $(-3)$ indeed
corresponds to a transition value, where the 
correlation function no longer decays algebraically at infinite 
Reynolds number.

Flows with the same inertial range spectrum may have distinct scaling 
functions, depending on their large-scale configuration; the non-universality 
of correlations can already be noticed by looking at the prefactor of
a first order expansion in $r/\xi$.
However, these functions commonly decay exponentially at distances
much larger than the correlation length. On the basis of
experimental results \cite{soap}, we argue that pair correlations 
between large-scale coherent vortices may be liquid-like.

Most of the statistical remarks made above are not qualitatively 
modified in three dimensions. It can be shown that for a Kolmogorov $-5/3$ law,
the vorticity correlations decay algebraically and satisfy
relation (\ref{scaling}) with $\eta=4/3$ \cite{monin}. In the general case,
at finite Reynolds numbers, the circulation modulus introduced in relation
(\ref{Omega}) is a sum of random variables if the integrated area extends 
over many integral scales, see Eq.(\ref{finsiz}). However, it scales 
anomalously in the inertial range, see Eq.(\ref{omall2}).
This is probably a feature satisfied by other global quantities 
in turbulent flows, 
like the power consumption studied experimentally in refs.\cite{nature,labbe}. 
The present analysis is consistent with some of the results presented in these
two references: in a closed flow confined to a 
size of order of one energy injection scale,
the power consumption fluctuates strongly. In turn,
in an open flow, many large eddies can develop
and the fluctuations are Gaussian,
according to the central limit theorem.

\acknowledgments{ The author is grateful to F. Lund for fruitful 
discussions and a critical reading of the manuscript. 
This work was supported by Fondecyt Grant 3970013 and a C\'atedra Presidencial 
en Ciencias.
}

\begin{table}
  \caption{Two-dimensional turbulence and critical phenomena compared.}
  \label{table}
\begin{tabular}{ll}
$2d$ turbulence &Critical phenomena\\
\tableline
$R_{\omega}(r)=\langle\vec{\omega }({\bf x})\cdot\vec{\omega }({\bf x}+{\bf r})
\rangle/\langle\vec{\omega }^2({\bf x})\rangle$
&Spin correlations $\langle S({\bf x})S({\bf x}+{\bf r})\rangle$\\
$1/Re\ (\rightarrow0)$ & $t=|T-T_c|/T_c$ $(\rightarrow0)$\\
$R_{\omega}(r)\sim(r/a)^{-\eta}f(r/l)$
&$\langle S({\bf x})S({\bf x}+{\bf r})\rangle\sim r^{2-d-\eta}f(r/\xi)$\\
Dissipation scale $a$ & Lattice spacing $a$\\
Integral length: $l\sim a(1/Re)^{-\nu}$ 
& Correlation length $\xi\sim  t^{-\nu}$\\
$\chi\equiv A^{-1} \left\langle\left(\int_A d{\bf x}\ 
\vec\omega({\bf x})\right)^2\right\rangle/
\langle\vec\omega^2\rangle\sim a^2(1/Re)^{-\gamma}$
& Susceptibility: $\chi\propto
\left\langle\left(\sum_i{\bf S}_i\right)^2\right\rangle\sim t^{-\gamma}$ 
\\
Exponents relation: $\gamma=\nu(2-\eta)$ & $\gamma=\nu(2-\eta)$\\
Kraichnan $\&$ Batchelor's theories: $\eta=0$, $\nu=1/2$ & Gaussian result: 
$\eta=0$, $\nu=1/2$\\
\tableline
\end{tabular}
\end{table}


\begin{thebibliography}{99}
\bibitem{note1} It can be shown that the scaling behaviors 
of $R_{\omega}$ are independent of the detailed shape of the dissipative 
spectrum, provided it is continuous in the case $\mu<3/2$.
\end{thebibliography}
\end{document}